\documentstyle[aps,prb,prl,epsfig,floats]{revtex}
\begin{document}
\draft
\twocolumn[\hsize\textwidth\columnwidth\hsize
\csname @twocolumnfalse\endcsname
\title{Possible isotope effect on the resonance peak formation 
in high-T$_c$ cuprates}
\author{I. Eremin$^{1,2}$, O. Kamaev$^{1}$, and M.V. Eremin$^1$}
\address{$^1$Physics Department, Kazan State University, 420008 Kazan, Russia}
\address{$^2$Institut f\"ur Theoretische Physik, Freie Universit\"at 
Berlin, D-14195 Berlin, Germany}
\date{\today}
\maketitle
\begin{abstract}
  Within effective $t-J$ Hamiltonian 
  we analyze the influence of electronic correlations and  
  electron-phonon interaction on the dynamical spin susceptibility in 
  layered cuprates. We find an isotope effect on the resonance peak 
  in the magnetic spin susceptibility, $\mbox{Im }\chi({\bf q},\omega)$, 
  seen by inelastic neutron scattering. It results from both  
  the electron-phonon coupling and the electronic correlation effects 
  taken into 
  account beyond random phase approximation(RPA) scheme. We find at optimal 
  doping the isotope coeffiecient $\alpha_{res} \approx 0.4$ 
  which can be further tested experimentally.              
\end{abstract}
] 

\narrowtext 
An understanding of the elementary and the spin excitations in
high-T$_{c}$ cuprates is of central significance. For example, 
it is known that the
Cooper-pairing scenario via the exchange of antiferromagnetic 
spin fluctuations was quite successful in explaining the 
various features of superconductivity in
hole-doped cuprates such as $d_{x^{2}-y^{2}}$-wave symmetry of the
superconducting order parameter and its feedback on the elementary and spin
excitatitions\cite{chubuk}. Most importanly, in this scenario the dynamical
spin susceptibility, $\chi ({\bf q},\omega )$, controls mainly the
superconducting and normal state properties of the layered cuprates
\cite{chubuk}. 
One of the key experimental fact in the phenomenology of high-T$_{c}$
cuprates is the occurrence of a so-called resonance peak in the inelastic
neutron scattering(INS) experiments\cite{he,bourges}. It occurs below T$_{c}$ in
the dynamical spin susceptibility, $\chi ({\bf q},\omega )$, at the
antiferromagnetic wave vector ${\bf Q}=(\pi ,\pi )$ and $\omega \approx 
\omega _{res}$
which is of the order of 40meV in the optimally doped cuprates. Its feedback
in various electronic properties like optical conductivity, Raman response 
function, and elementary excitations 
has been observed experimentally by various techniques\cite{chubuk}. 
Furthermore, its successful explanation within
spin-fluctuation-mediated Cooper-pairing together with 
$d_{x^{2}-y^{2}}$-wave symmetry of the superconducting order parameter favors
this scenario as a basic one for superconductivity in the cuprates. 
On the other hand, recent experiments indicate that also
electron-phonon interaction influences strongly their 
behavior\cite{zhao00,khasan,astuto,pintsch,kulic}.
In particular, the observation of the relatively large isotope effect in
various characteristics of cuprates like penetration depth\cite{khasan},
'kink'-structure seen by ARPES\cite{lanzara} still raises a question: what is
the role of phonons in determining the superconducting properties of
cuprates? 

Here, we derive an effective 
$t-J$ Hamiltonian where both the hopping integral, $t$, 
and the superexchange
interaction between neighboring spins, $J$, are renormalized by phonons. 
We analyze the influence of the electron-phonon interaction on the dynamical
spin susceptibility in layered cuprates.
In particular, we find an isotope
effect on the resonance peak in the magnetic spin susceptibility, $\mbox{Im}
\chi ({\bf q},\omega )$. It results
from both the electron-phonon coupling and the electronic correlation
effects taken into account beyond random phase approximation(RPA) scheme. 
We show that even if the
superconductivity is driven by the magnetic exchange the characteristic
energy features of cuprates can be significantly renormalized by the strong 
electron-phonon interaction.

{\it Effective Hamiltonian}  We start from the atomic limit of the
three-band $p-d$ Hamiltonian 
\begin{eqnarray}
H_{0} &=&\sum \epsilon _{d}d_{i\sigma }^{\dagger }d_{i\sigma }+\sum \epsilon
_{p}p_{i\sigma }^{\dagger }p_{i\sigma }+\sum U_{d}n_{d\uparrow
}n_{d\downarrow }  \nonumber \\
&&+\sum U_{p}n_{p\uparrow }n_{p\downarrow }+\sum \hbar \omega _{{\bf q}}f_{%
{\bf q}}^{\dagger }f_{{\bf q}}  \label{eq1}
\end{eqnarray}%
where $\epsilon _{d}$ and $\epsilon _{p}$ are the on-site energies of the
copper and the oxygen holes, $n_{d\sigma }=d_{i\sigma }^{\dagger }d_{i\sigma }$
and $n_{p\sigma }=p_{i\sigma }^{\dagger }p_{i\sigma }$ are the copper $3d$
and oxygen $2p$ hole densities for site $i$, respectively. $U_{d}$ and $U_{p}
$ refer to the on-site copper and oxygen Coulomb repulsion, respectively. $f_{%
{\bf q}}^{\dagger }$ denotes the phonon creation operator and $\hbar
\omega _{\bf q}$ is a phonon energy dispersion. The hopping
term between copper and oxygen 
\begin{eqnarray}
H_{2}=\sum_{\sigma }t_{pd}\left( d_{a\sigma }^{\dagger }p_{b\sigma
}+h.c.\right) \quad,
\label{h1}
\end{eqnarray}
and the electron-phonon interaction
\begin{eqnarray}
H_1 & = & 
\sum_{l=d,p}g_{l}n_{l}\left( f_{{\bf -q}}^{\dagger }+f_{{\bf q} }\right) 
\quad,
\label{h2}
\end{eqnarray}
we consider as a perturbation. 
Here, $t_{pd}$ is a hopping term between copper and oxygen, $g_{l}$ is a 
electron-phonon coupling strength at the site $l$. This notation is similar
to the simplified Holstein model where the migrating charge 
interacts locally with breathing phonon modes 
forming electron-vibrational states.

To derive an effective $t-J$ Hamiltonian we employ the canonical
Schrieffer-Wolf-like transformations $e^{-S}He^{S}$ \cite{wolf,kugel}. The
matrix of the unitary transformation for the initial Hamiltonian is found by
excluding the odd terms with respect to the hopping integral with an accuracy
up to the sixth order perturbation theory. Then the $S$-operator consists
of the sum of five terms. Each of them is determined by the following
iteration procedure 
\begin{eqnarray}
\left[ H_{0}S_{1}\right]  &=&-H_{2},\quad \left[ H_{0}S_{2}\right] =-\left[
H_{1}S_{1}\right] ,  \nonumber \\
\left[ H_{0}S_{3}\right]  &=&-\left[ H_{1}S_{2}\right] -\frac{1}{3}\left[ %
\left[ H_{2}S_{1}\right] S_{1}\right] ,  \nonumber \\
\left[ H_{0}S_{4}\right]  &=&-\left[ H_{1}S_{3}\right] -\frac{1}{3}\left[ %
\left[ H_{2}S_{1}\right] S_{2}\right] -\frac{1}{3}\left[ \left[ H_{2}S_{2}%
\right] S_{1}\right] ,  \nonumber \\
\left[ H_{0}S_{5}\right]  &=&-\left[ H_{1}S_{4}\right] -\frac{1}{3}\left[ %
\left[ H_{2}S_{1}\right] S_{3}\right] -\frac{1}{3}\left[ \left[ H_{2}S_{3}%
\right] S_{1}\right]   \nonumber \\
&&-\frac{1}{3}\left[ \left[ H_{2}S_{2}\right] S_{2}\right] +\frac{1}{45}%
\left[ \left[ \left[ \left[ H_{2}S_{1}\right] S_{1}\right] S_{1}\right] S_{1}%
\right] .  \label{s}
\end{eqnarray}
The calculations are straightforward and their details will be given
elsewhere. Note, in the second order perturbation 
the effective hopping integral, $t_{ij}$, appears. 
It is further renormalized by the electron-phonon interaction in the fourth
order term where we introduce the average over the phonons. 
Similarly, the superexchange
interaction occurs in the fourth order perturbation theory and  
its renormalization 
takes place in the sixth order term. Finally, the relevant effective
Hamiltonian is given by 
\begin{equation}
H=\sum_{ij}t_{ij}\Psi _{i}^{pd,\sigma }\Psi _{j}^{\sigma
,pd}+\sum_{i>j}J_{ij}\left[ ({\bf S}_{i}{\bf S}_{j})-\frac{n_{i}n_{j}}{4}%
\right] \quad .  
\label{t-J}
\end{equation}
Note, in general case the effective Hamiltonian contains also the Coulomb
interaction between doped holes and the interaction of quasiparticles via
the phonon field. 
We dropped these terms here, because they do not contribute directly 
to the spin susceptibility. 
In Eq.(\ref{t-J}) we use the projecting Hubbard-like operators 
$\Psi_i^{\alpha,\beta}=\mid i,\alpha ><i,\beta \mid $ in order to satisfy no 
double occupancy constraint. 
The index $pd$ corresponds to a Zhang-Rice singlet 
formation with one hole placed on the copper site whereas the second hole 
is distributed on the neighboring oxygen sites \cite{rice}. 
Note, 
$t_{ij}=t_{ij}^{0}e^{-\gamma E_{i}^{\ast }/\hbar \omega _{i}^{\ast }}$
where $t_{ij}^{0}$ is the effective hopping integral without taking into
account electron-phonon interaction, $E_{i}=(g_{i}^{\ast })^{2}/\hbar
\omega _{i}^{\ast }$ is the so-called polaron stabilization energy of the 
copper-oxygen singlet state and $0<\gamma<1$. Note, 
from the experimental data\cite{zhao1} the whole exponential factor 
was estimated to be $\gamma E_{i}^{\ast }/\hbar \omega _{i}^{\ast } 
\approx 0.92$ around the optimal doping and its value is increasing 
upon decreasing doping. We further assume that the lifetime of the Zhang-Rice 
singlet is much larger than the relaxation time of the local deformations.

Similarly the superexchange interaction between
nearest copper spins is given by 
\begin{eqnarray}
J & = &J_{0}\left\{ 1+\frac{3\hbar }{\Delta _{pd}^{2}}\left[ E_{p}\omega
_{p}\coth \left( \frac{\hbar \omega _{p}}{2k_{B}T}\right) \right. \right.  
\nonumber \\
&&\left. \left. +E_{d}\omega _{d}\coth \left( \frac{\hbar \omega _{d}}{
2k_{B}T}\right) \right] \right\}   \quad,
\label{J}
\end{eqnarray}
where $\Delta_{pd}=\epsilon_{p}-\epsilon_{d}+U_{p}-U_{d}$ is the energy
transfer from copper to oxygen and is known to be of the order of 
$1.5$eV in the cuprates. Here, $J_{0}$ is the superexchange
interaction of copper spins via the intermediate oxygen atom in the absence
of phonons. We took the phonon frequency of the order of $\omega ^{\ast
}\approx \omega_{p}\approx \omega _{d}=$50meV which roughly corresponds to the
energy of the longitudinal optical(LO) bond stretching phonon mode in
cuprates. According to the recent experiments\cite{astuto,pintsch} it
may play an essential role in the physics of cuprates. The so-called
polaronic stabilization energy $E^{\ast }\approx E_{p}\approx E_{d}$ was
estimated of the order of 0.5eV in accordance with the measurements of the
isotope effect in cuprates\cite{zhao}.

{\it Dynamical spin susceptibility} To derive the dynamical spin
susceptibility in the superconducting state we use the method suggested by
Hubbard and Jain\cite{hubjain} that allows to take into account strong
electronic correlations. First we add the external magnetic field 
applied along $c$-axis into the effective Hamiltonian  
\begin{equation}
H_i=\mbox{Re} \sum_{\bf q} h_{\bf -q}e^{i(\omega t-{\bf qR_i})} +h_{\bf q}e^{-i(\omega t - {\bf qR_i})}
\quad.
\label{magnfeld}
\end{equation}
Then we write an equation of motion for the $\Psi$ operators using the
Roth-type of the decoupling scheme\cite{plakida} and expanding the $%
P^{\sigma}_{pd} = \left\{\Psi_i^{\uparrow,pd} \Psi_i^{pd, \uparrow} \right\} = 
\frac{1+\delta_i}{2} + \sigma \mbox{Re} \sum_{{\bf q}} \left[ S_{{\bf -q}}^z
e^{-i({\bf q R_j} - \omega t)}+ S_{{\bf q}}^z e^{i({\bf qR_j} - \omega t)} %
\right]$ up to the first order in 
$S_{{\bf q}}= \chi^{zz}({\bf q}, \omega) h_{\bf q}$. In particular, 
\begin{eqnarray}
\lefteqn{i \hbar \frac{\partial \Psi_{\bf k}^{ -\sigma, pd}}{\partial t} =
(\epsilon_{\bf k}-\mu) \Psi_{\bf k}^{- \sigma, pd} + \Delta_{\bf k} \Psi_{\bf
-k}^{pd, \sigma} }  \nonumber \\
&& + \left[ \left(\frac{J_{{\bf q}}}{2} - t_{{\bf k-q}} \right) S_{{\bf q}} -
\frac{h_{\bf q}}{2}\right] \Psi_{{\bf k-q}}^{-\sigma, pd} e^{-i\omega t}  \nonumber \\
&& + \left[ \left(\frac{J_{{\bf - q}}}{2} - t_{{\bf k+q}} \right) S_{{\bf - q
}} -\frac{h_{\bf -q}}{2}\right] \Psi_{{\bf k+q}}^{-\sigma, pd} e^{i\omega t} \quad, 
%\nonumber
%\\
%%
%&& i \hbar \frac{\partial \Psi_{{\bf - k}}^{pd, \sigma}}{\partial t} = -
%(\epsilon_{{\bf k}}-\mu) \Psi_{{\bf - k}}^{pd, \sigma} + \Delta^*_{{\bf k}} \Psi_{{\bf -k}}^{-\sigma, pd}  \nonumber \\
%&& + \left[ \left(\frac{J_{{\bf - q}}}{2} - t_{{\bf -k-q}} \right) S_{{\bf -
%q}} -\frac{h_{\bf -q}}{2}\right] \Psi_{{\bf -k-q}}^{pd \sigma} e^{-i\omega t}  \nonumber
%\\
%&& + \left[ \left(\frac{J_{{\bf q}}}{2} - t_{{\bf -k+q}} \right) S_{{\bf q}}
%-\frac{h_{\bf q}}{2}\right] \Psi_{{\bf -k+q}}^{-\sigma pd} e^{i\omega t} \quad.  
\label{eqm}
\end{eqnarray}
and the similar expression occurs for $\Psi_{{\bf - k}}^{pd, \sigma}$.
Here, $\Delta_{{\bf k}} = \frac{\Delta_0}{2} \left(\cos k_x - \cos
k_y\right) $ is $d_{x^2-y^2}$-wave superconducting gap, $J_{{\bf q}} = J
\left(\cos k_x + \cos k_{y} \right)$ is the Fourier transform of the
superexchange interaction on a square lattice.

The expression for the longitudinal component of the dynamical spin
susceptibility can be obtained from the relation 
\begin{equation}
\langle \Psi _{i}^{pd,\uparrow }\Psi _{i}^{\uparrow, pd}\rangle -\langle \Psi
_{i}^{pd,\downarrow }\Psi _{i}^{\downarrow, pd}\rangle =0  \quad,
\label{forsus}
\end{equation}
and using the Bogolyubov-like transformations to the new quasiparticle
states 
\begin{eqnarray}
X_{{\bf k}}^{-\tilde{\sigma},pd} &=&u_{{\bf k}}\Psi _{{\bf k}}^{-\sigma,
pd}+v_{{\bf k}}\Psi _{{\bf -k}}^{pd,\sigma }\quad ,  \nonumber \\
X_{{\bf -k}}^{pd,\tilde{\sigma}} &=&u_{{\bf k}}\Psi _{{\bf k}}^{pd,\sigma }-v_{%
{\bf k}}\Psi _{{\bf -k}}^{-\sigma, pd}\quad .  \label{bogo}
\end{eqnarray}%
Here, $u_{{\bf k}}^{2}=\frac{1}{2}\left( 1+\frac{\epsilon _{{\bf k}}-\mu }{%
E_{{\bf K}}}\right) $ and $v_{{\bf k}}^{2}=\frac{1}{2}\left( 1-\frac{%
\epsilon _{{\bf k}}-\mu }{E_{{\bf K}}}\right) $ are the Bogolyubov
coefficients, $\mu$ is a chemical potential, 
and $E_{{\bf k}}=\sqrt{(\epsilon _{{\bf k}}-\mu )^{2}+\Delta _{
{\bf k}}^{2}}$ is the energy dispersion in the superconducting state.
Substituting Eq.(\ref{bogo}) in Eq.(\ref{forsus}) and using the equations of
motion (\ref{eqm}) 
one obtains the expression for the dynamical spin susceptibility in
the form 
\begin{equation}
\chi ({\bf q},\omega )=\frac{\chi _{0}({\bf q},\omega )}{J_{{\bf q}}\chi
_{0}({\bf q},\omega )+\Pi ({\bf q},\omega )+Z({\bf q},\omega )}  \quad.
\label{chi}
\end{equation}
This is a central result of our paper. Here, $\chi _{0}({\bf q},\omega )$
is the usual BCS-like Lindhard response function,  $\Pi ({\bf q},\omega )$
and $Z({\bf q},\omega )$ result from the strong electronic 
correlation effects. In the
normal state the expression for  $\Pi ({\bf q},\omega )$ has been
obtained by Hubbard and Jain \cite{hubjain}. 
In the superconducting state it is given by
\begin{eqnarray}
\lefteqn{\Pi ({\bf q},\omega )=\frac{P_{pd}}{N}\sum_{\bf k}}  \nonumber \\
&&u_{{\bf k}}u_{{\bf k+q}}\left( u_{{\bf k}}u_{{\bf k+q}}+v_{{\bf k}}v_{{\bf %
k+q}}\right) \frac{t_{{\bf k}}f_{{\bf k}}-t_{{\bf k+q}}f_{{\bf k+q}}}{\omega
+i0^{+}+E_{{\bf k}}-E_{{\bf k+q}}}  \nonumber \\
&&+v_{{\bf k}}v_{{\bf k+q}}\left( v_{{\bf k}}v_{{\bf k+q}}+u_{{\bf k}}u_{%
{\bf k+q}}\right) \frac{t_{{\bf k}}(1-f_{{\bf k}})-t_{{\bf k+q}}(1-f_{{\bf %
k+q}})}{\omega +i0^{+}-E_{{\bf k}}+E_{{\bf k+q}}}  \nonumber \\
&&+u_{{\bf k}}v_{{\bf k+q}}\left( u_{{\bf k}}v_{{\bf k+q}}-u_{{\bf k+q}}v_{%
{\bf k}}\right) \frac{t_{{\bf k}}f_{{\bf k}}-t_{{\bf k+q}}(1-f_{{\bf k+q}})}{%
\omega +i0^{+}+E_{{\bf k}}+E_{{\bf k+q}}}  \nonumber \\
&&+u_{{\bf k+q}}v_{{\bf k}}\left( v_{{\bf k}}u_{{\bf k+q}}-u_{{\bf k}}v_{%
{\bf k+q}}\right) \frac{t_{{\bf k}}(1-f_{{\bf k}})-t_{{\bf k+q}}f_{{\bf k+q}}%
}{\omega +i0^{+}-E_{{\bf k}}-E_{{\bf k+q}}}.  
\label{Pi}
\end{eqnarray}
The function $Z({\bf q},\omega)$ is written as follows
\begin{equation}
Z({\bf q},\omega )=\frac{1}{N}\sum_{{\bf k}}\frac{\omega +i0^{+}}{\omega
+i0^{+}+\epsilon _{{\bf k}}^{(1)}-\epsilon _{{\bf k+q}}^{(1)}} \quad.
\end{equation}
Here, $f_{\bf k}$ is the Fermi distribution function, 
$\epsilon^{(1)}_{\bf k} = \frac{1-\delta}{2} t_{\bf k}$, 
$\epsilon_{\bf k} = P_{pd} t_{\bf k}$ is the energy dispersion in the 
normal state, and $t_{\bf k} = 2 
t (\cos k_x + \cos k_y) + 4t'\cos k_x \cos k_y + 2t'' (\cos 2k_x + 
\cos 2k_y)$ is the Fourier transform of the hopping integral on a 
square lattice including nearest, next- and next-next-nearest neighbor 
hopping, respectively. 
The origin of the terms $\Pi({\bf q},\omega)$ and $Z({\bf q},\omega)$ 
relates to the no double occupancy constraint.
In particular, for the Coulomb repulsion $U=\infty$ and $J=0$ 
the dynamical spin susceptibility does not reduce to the standard 
Lindhard response function but is renormalized by the electronic correlation 
effects\cite{remark2}. 
For the $\Delta_{\bf k} =0$ Eq.(\ref{chi}) agrees with the normal 
state result for the dynamical spin 
susceptibility\cite{hubjain,zavidon,eremin}.

{\it Results and Discussion} Inelastic neutron scattering (INS) measurements
probe directly the imaginary 
%
%\vspace*{-1.0cm}
\begin{figure}[t]
\centerline{\epsfig{clip=,file=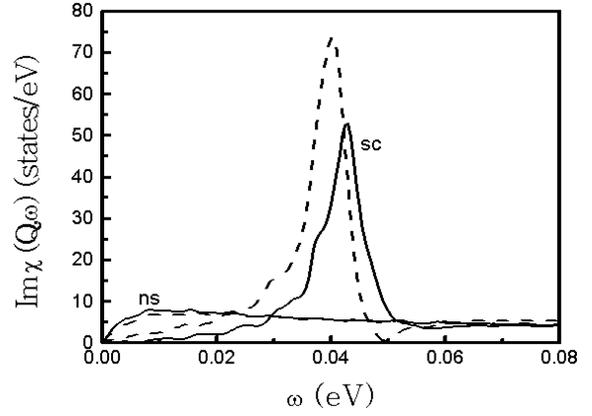,width=7.7cm,angle=0}}
\caption{Calculated imaginary part of the dynamical spin susceptibility 
Im$\chi ({\bf Q},\protect\omega)$ using Eq.(\protect\ref{chi}) and 
$J$=0.3$t$ in the normal and superconducting state at optimal doping. 
Here, we use the superconducting gap 
$\Delta_0 = 0.14t$ (28 meV) and $T_c \approx 0.04t$ (90 K)  
for optimal doping from our earlier
calculations of the mean-field phase diagram \protect\cite{prbearlier}. 
To illustrate the role
of  $Z({\bf q},\omega )$  we also show the results for 
$Z({\bf q} ,\omega )=0$ (dashed curve). Note, the damping was chosen 
$\Gamma = 1.5$meV. 
}
\label{fig1}
\end{figure}
part of the dynamical spin susceptibility. Therefore, it is of interest to
analyze the role played by the electronic correlations on
the 'resonance' peak formation seen by INS\cite{bourges}. 
Using various approaches this feature was well
understood mainly as a result of the spin
density wave (SDW) collective mode formation at $\omega = \omega_{res}$, 
{\it i.e.} when the denominator of the  spin susceptibility at the
antiferromagnetic wave vector ${\bf Q}$ is close to zero \cite{many,norman}.

\begin{figure}[t]
\centerline{\epsfig{clip=,file=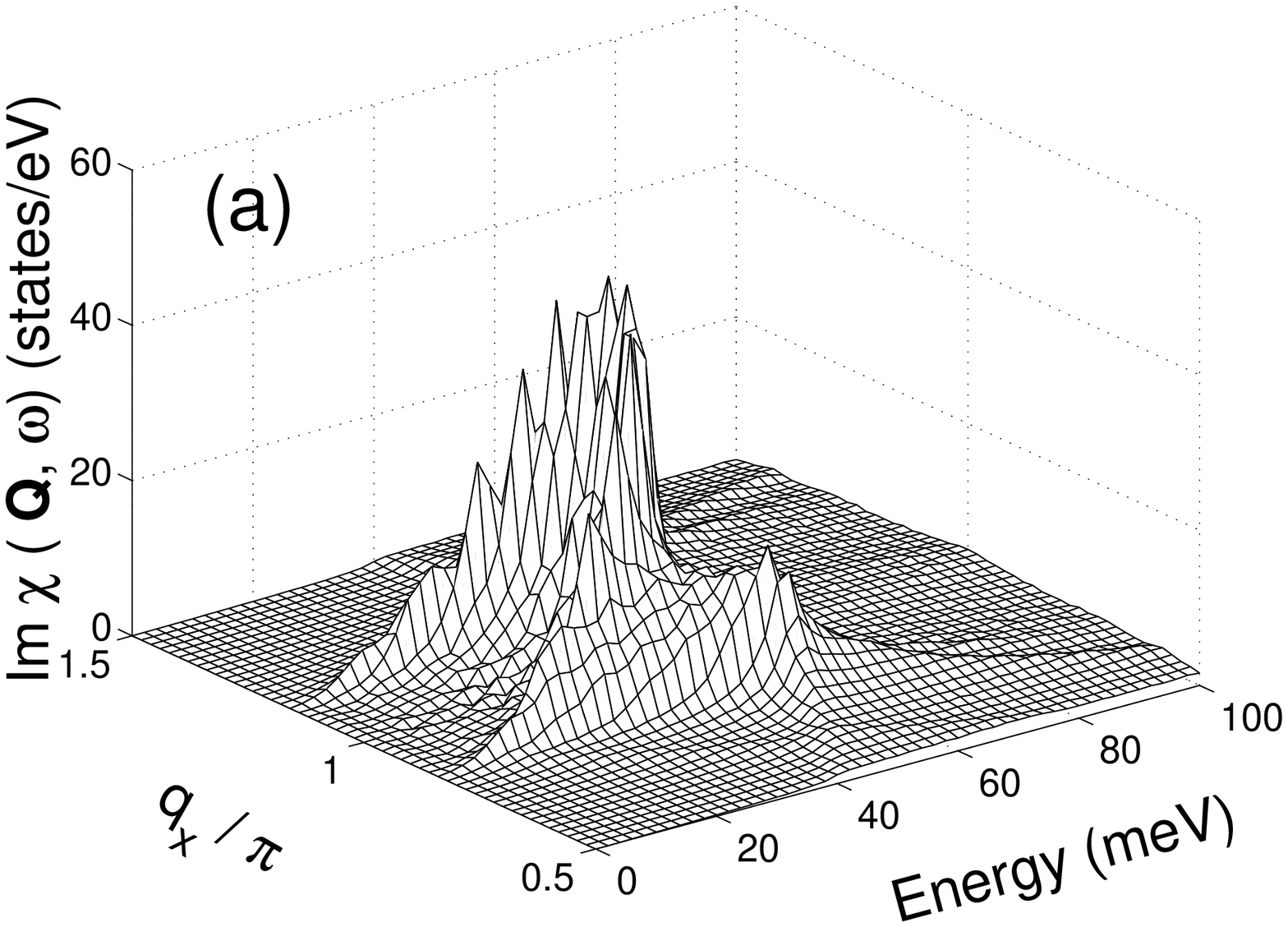,width=7.9cm,angle=0}}
\centerline{\epsfig{clip=,file=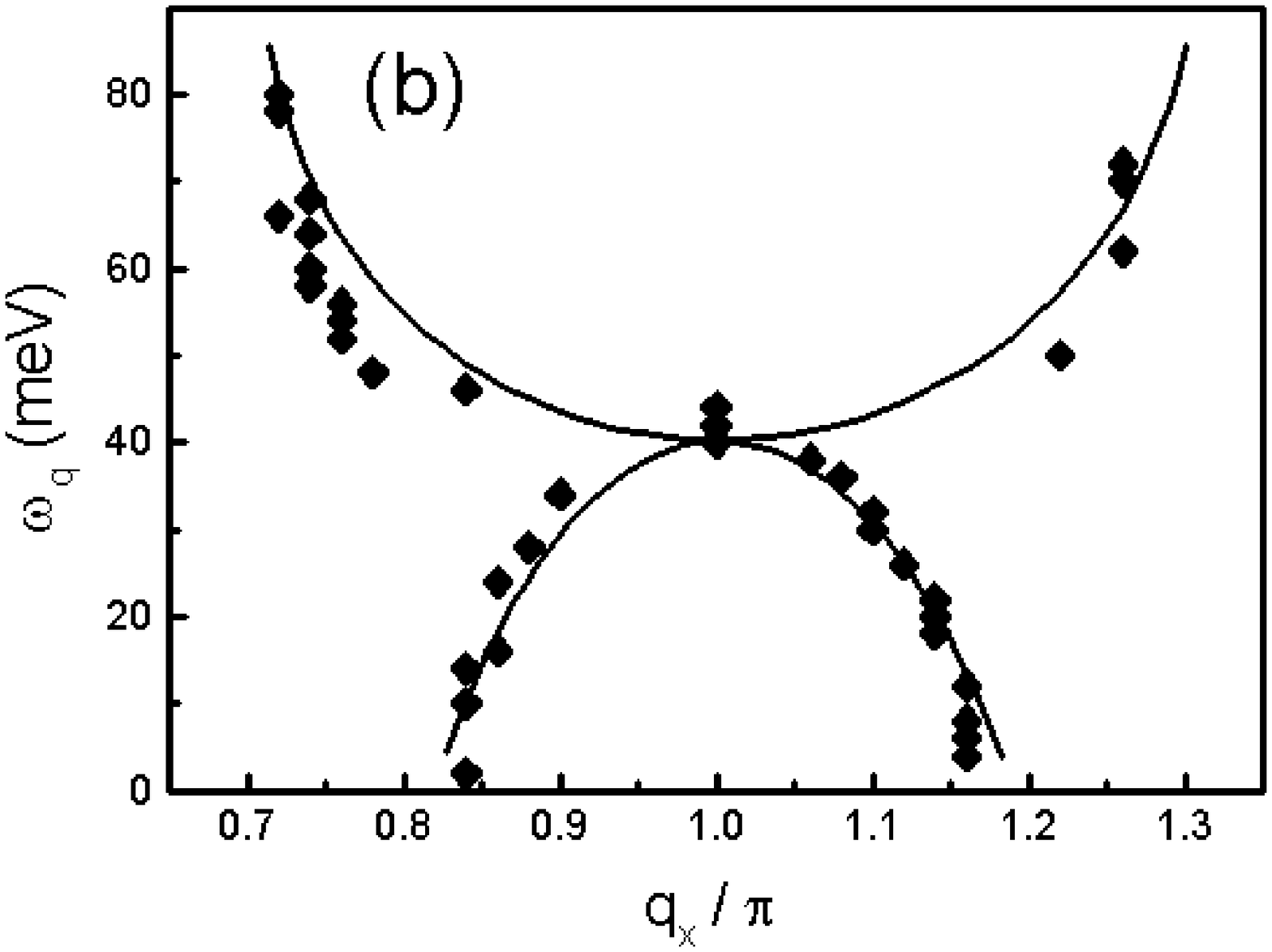,width=7.9cm,angle=0}}
\caption{ (a) Calculated frequency and momentum dependence of  
Im$\chi ({\bf q},\omega )$ away from ${\bf Q}=(\protect\pi ,\protect\pi)$. 
(b) The dispersion of the resonance peak.
Two branches of the dispersion curves are in good agreement with recent
experimental data \protect\cite{arai,bourge2}. Note, the solid curve is a 
guide to the eye.}
\label{fig2}
\end{figure}

In Fig. \ref{fig1} we show the results of our calculations for the Im $\chi (%
{\bf Q},\omega )$ as a function of frequency in the normal and the
superconducting state. Here, we use $t=200$meV, $t^{\prime}=-0.1t$, and 
$t^{\prime\prime}=0.02t$ at optimal doping.
Clearly in the normal state the spin
fluctuation spectrum is characterized by a broad feature which starts 
around 10meV and extends up to a higher frequencies. 
In the superconducting state it strongly renormalizes due to a presence 
of the $d_{x^2-y^2}$-wave gap ($\Delta_{\bf k} = \frac{\Delta_0}{2}
(\cos k_x -\cos k_y)$) that leads to a resonance peak formation 
similar to the RPA result\cite{many,norman}. However, due to a strong 
frequency dependence  
of $Z({\bf Q},\omega)$ and $\Pi ({\bf Q},\omega)$ the spectral weight of the 
resonance peak is redistributed away from $(\pi,\pi)$ (see the dashed curve 
for comparison) leading to a well-pronounced dispersion of the latter. 
This is
illustrated in Fig.\ref{fig2}(a) where we show the calculated frequency and 
momentum dependence of Im $\chi({\bf q},\omega)$ away from the 
antiferromagnetic wave vector ${\bf Q} = (\pi,\pi)$. The dispersion of the 
resonance excitations is clearly visible and is shown in Fig.\ref{fig2}(b) 
as a function of $q_x$ ($q_y = \pi$). As one sees there are
well-pronounced dispersion curves $\propto {\bf q}^{2}$ in good agreement
with 
%
%\vspace*{-1.5cm}
experiment\cite{arai,bourge2}. Note, that in the RPA the dispersion
of the resonance is much weaker (in particular the upper branch of 
the dispersion) due to a $\delta $-function character of the
resonance condition\cite{norman} while 
$Z-$ and $\Pi$-terms, in particular, lead to a redistribution of the spectral 
weight away from ${\bf Q}=(\pi ,\pi )$. Note, due to the tetragonal symmetry 
the same dispersion takes place for $q_y$ ($q_x = \pi$).

%
%\vspace*{-1.5cm}
\begin{figure}[t]
\centerline{\epsfig{clip=,file=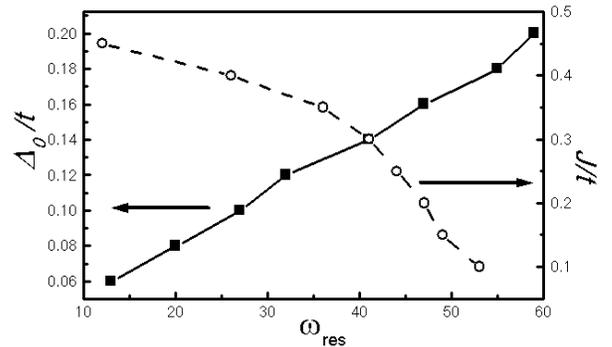,width=7.9cm,angle=0}}
\caption{ Calculated dependence of the resonance frequency $\omega_{res}$ 
on the superexchange coupling constant $J$ (open circles) 
and on the value of the 
$d_{x^2-y^2}$-wave superconducting gap $\Delta_0$ (filled squares). Note, the 
lines are the guides to the eye.   
}
\label{fignew}
\end{figure}
The position of the resonance peak is determined mainly by the magnitude
of the $d_{x^2-y^2}$-wave superconducting gap, $\Delta_0$, 
the superexchange coupling 
constant, $J$, and by the proximity of the Fermi energy to the extended 
saddle point (the so-called van-Hove 
singularity) in the density of states determined by the ratio of $t'/t$.  
In order to illustrate this dependence we show in Fig. \ref{fignew} 
the resonance peak position as a function of $J/t$ and $\Delta_0 / t$ at 
optimal doping.   
One can clearly see that it depends almost linearly 
on $2\Delta_0$. This follows from our Eq. (\ref{chi}). In particular, 
the superconducting gap determines mainly the 
position of the continuum of the spin excitations from 
Im$\chi_0$ and which is around 
2$\Delta_0$\cite{many}. Thus, at the fixed value of $J/t$ and $t'/t$ 
this continuum influences also the resonant condition because the difference 
between the position of the resonance peak (SDW collective mode) and 
the continuum of the states has to remain the same\cite{many}. 
On the other hand, 
for the fixed values of $t'/t$ and $\Delta_0/t$ the resonance 
condition depends mainly on the value 
of $J$. Furthermore, its dependence is more complicated than in the case of 
$\Delta_0$. At the relatively small values of the superexchange coupling 
the resonance lies close to the continuum and is more 
sensitive to the change of the superconducting gap magnitude 
than to the superexchange coupling constant. 
On the contrary, one could see from the Fig. \ref{fignew} 
that if the resonance peak is at small energies and lies relatively far 
from the continuum (which happens for large values of $J/t$) it  
will be most influenced by the change of $J/t$ values.
We note that the shift of the resonance peak to the lower frequency enhances 
its intensity and visa versa. 
In addition, the ratio of $t'/t$ which determines the nesting of the 
Fermi surface and the position of van-Hove singularity 
influences also the resonance peak. Its influence is somewhat 
similar to $J$ with one important difference. In particular, an 
increase of $t'/t$ rather weakens the intensity of the resonance peak 
than changes the position itself (not shown). Note, that our analysis 
agrees qualitatively with the previous ones\cite{many}.
Most importantly we find that by the slight 
variation of all parameters one 
could find another realistic set of $\Delta_0/t$, $J/t$, and $t'/t$ 
which would also correctly explain the position of the resonance peak at 
optimal doping\cite{revremark}. 
This means that at present stage we could make only a 
qualitative analysis of the experimental data. Further factors like 
orthorhombic distortions may influence 
the difference between the resonance peak intensity in 
Bi$_2$Sr$_2$CaCu$_3$O$_{8+x}$ (BSCCO) and YBa$_2$Cu$_3$O$_7$(YBCO). 

Finally, we discuss the influence of the electron-phonon interaction on the
resonance peak formation by changing the isotope mass of $
^{16}$O by $^{18}$O. This shifts the average
frequency of the LO phonon mode and consequently renormalizes 
the hopping integral $t$ and 
the superexchange coupling constant, $J$. 
Most importantly, the electron-phonon interaction changes most 
dramatically the 
hopping integral $t$ rather than the superexchange coupling $J$.
In particular, as can be seen from Eq.(\ref{J}) the superexchange coupling 
constant $J$ changes less than 1$\%$ upon substituting the 
isotopes\cite{jetpl} which agrees well with experimental data\cite{morris0}. 
Therefore, there is almost no influence 
of the isotope substitution 
on the resonance peak determined from RPA, since in this approximation 
its formation 
is determined mainly by $J$. In particular, we find {\it within RPA} 
no change in the $\omega_{res}$ value upon changing the isotopes.
In the case of Eq.(\ref{chi}) the most important contribution to 
the isotope effect on the resonance peak appears due to $\Pi ({\bf q}
,\omega )\propto t_{k}$. In particular, using our estimation given above 
we find that at optimal doping the hopping integral changes by 6$\%$ upon
replacing $^{16}$O by $^{18}$O. This results in the lowering of the resonance 
frequency at ($\pi,\pi$) from 41meV for the $^{16}$O isotope  towards 
39meV for the $^{18}$O sample. This leads to $\alpha _{res}= 
- \frac{d\ln
\omega _{res}}{d\ln M}\approx 0.4$ for optimally-doped cuprates.
This effect is beyond the experimental error and can be further tested 
experimentally. Furthermore, in the underdoped 
cuprates one may expect larger isotope effect due to a larger value of 
 $\gamma E_i^{\ast}/\hbar \omega^{\ast}_{i}$\cite{zhao1}.
At the same time the superconducting transition temperature which is
determined by $J$ shows much weaker isotope effect and is around $\alpha
_{T_c}\approx $0.05 \cite{jetpl}. Therefore, even if the
superconductivity is driven by the magnetic exchange the resonance peak 
formation can be significantly renormalized by the strong 
electron-phonon interaction.

To summarize, we analyze the influence of the electronic correlations and
the electron-phonon interaction on the dynamical spin susceptibility in
layered cuprates. The electronic correlations taken beyond RPA redistribute the
spectral weight 
of the resonance peak away from ($\pi,\pi$) 
leading to the pronounced dispersion. This is in good agreement with recent 
INS data\cite{arai,bourge2}. Furthermore, we find the isotope
effect on the resonance peak due to strong coupling of the carriers to 
LO phonon mode. It results from both 
electron-phonon coupling and electronic correlation effects. In contrast
to the small isotope effect on the superconducting transition temperature we
find larger isotope coefficient on the resonance peak  
$\alpha _{res}\approx 0.4$ in optimally-doped cuprates. We also would like 
to note that the value of the isotope coefficient depends strongly on the 
value of the exponential factor. Therefore, 
the experimental verification of our prediction is desirable.  
In particular, 
it would put a strong constraint on the ingredients 
the theory of cuprates must contain. 

It's pleasure to thank P. Bourges, A. Lanzara, T. Timusk, D. Manske for 
useful discussions and M. Mali for critical reading of the manuscript. 
The work of I.E. is supported by INTAS grant No. 01-0654. 
M.V.E. and O.K. are supported by  the RFBR Grant No. 03-02-16550 and
RSP "Superconductivity" 98014-3.

\end{document}